\documentclass[%
 reprint,
 amsmath,amssymb,
 aps,
]{revtex4-1}

\usepackage{graphicx}
\usepackage{dcolumn}
\usepackage{bm}

\begin{document}

\preprint{APS/123-QED}

\title{Non-monotonic dynamics in the onset of frictional slip}

\author{Kasra Farain and Daniel Bonn}

\affiliation{Van der Waals–Zeeman Institute, Institute of Physics, University of Amsterdam, Science Park 904,  1098XH Amsterdam, The Netherlands\\
 k.farain@uva.nl\\d.bonn@uva.nl}

\begin{abstract}
The transition from static to dynamic friction is often described as a fracture-like instantaneous slip. However, studies on slow sliding processes aimed at understanding frictional instabilities and earthquakes report slow friction transients that are usually explained by empirical rate-and-state formulations. We perform very slow ($\sim nm/s$) macroscopic-scale sliding experiments and show that the onset of frictional slip is governed by continuous non-monotonic dynamics originating from a competition between contact aging and shear-induced rejuvenation. This allows to describe both our non-monotonic dynamics and the simpler rate-and-state transients with a single evolution equation.

\end{abstract}

\pacs{Valid PACS appear here}
\maketitle



More than two centuries ago, Amontons and Coulomb studied the motion of a block on a surface and concluded that sliding starts when the applied tangential force exceeds $\mu_{s}F_{N}$, where $\mu_{s}$ is the static friction coefficient and $F_{N}$ the force normal to the surface, determined by the weight of the block. Once in motion, the usual observation is that the frictional resistance drops to a smaller value given by the dynamic friction coefficient $\mu_{d}$. $\mu_{s}$ and $\mu_{d}$ are taken to be material properties that do not depend on block size or shape, as Da Vinci already noted \cite{1, 2}. These observations were later rationalized by Bowden and Tabor, who suggested that surface roughness leads to a linear proportionality between the real area of contact and the normal force. One then just needs to suppose that the interfacial shear strength of solids is a (material-dependent) constant to retrieve a constant friction coefficient \cite{1, 3}. However, reality is much more complicated. The static friction coefficient $\mu_{s}$ has been reported to vary considerably in the same system \cite{4, 5, 6}. It notably depends on the time that the surfaces have spent in contact. In the simplest scenario with two solids in static contact under a constant load, $\mu_{s}$ grows logarithmically in time \cite{7, 8, 9}. In line with Bowden and Tabor’s argument, such aging is usually attributed to a logarithmic increase of the area of real contact between the surfaces \cite{9, 10, 11}. The increase in contact area then also explains why $\mu_{s}>\mu_{d}$: a static interface causes more friction because it has aged and has a higher contact area, while, at the onset of slipping, the micro contacts break up, causing the total contact area and hence friction coefficient to shrink.

This simple picture, however, does not hold generically: the static friction coefficient $\mu_{s}$ and the area of real contact can even evolve in opposite directions \cite{12,13}. In addition, recent precise measurements of the real contact area show that although the dynamic friction is smaller than static friction, slipping can cause the area of real contact to grow instead of shrink \cite{14}. This clearly indicates that the idea of constant friction per unit contact area (shear strength) cannot account for the observed differences between $\mu_{s}$ and $\mu_{d}$. What has to be considered is the collective dynamics of the rough interface as a complex ensemble of many contact points within which the basic interactions occur. The most evident manifestation of such collective dynamics is the frictional aging at static interfaces. Yet, this universally observed phenomenon is not included in fracture-type arguments \cite{4,15, 16, 17, 18, 19} that reduce the static friction to some type of (chemical, adhesive) bonding between the surfaces.

To inspect the slip dynamics, we impose frictional motion on a small, rough interface between a mm-sized sphere and a flat smooth surface. This interfacial system contains a large distribution of individual asperities \cite{14, 20, 21}. The sliding experiments are performed using a rheometer (Anton Paar MCR302) that can apply very small angular velocities to a rheometer tool, a plate that is normally used to confine a liquid for rheology experiments.  A polytetrafluoroethylene (PTFE) sphere of 3.18 mm diameter and approximately 1.5 $\mu$m root-mean-square (rms) roughness (after removing the macroscopic spherical curvature) is attached off-center by a custom–made holder. We also use polypropylene spheres of 2.45 mm diameter and 1.3 $\mu$m rms-roughness. The surface topography of the spheres is measured with a Keyence optical profilometer. By lowering the rheometer head, the sphere touches a smooth ($<1$ nm rms-roughness) and clean glass coverslip to make a Hertz contact circle with a typical normal force (measured by the rheometer) of less than 100 mN. Subsequently, the rheometer tool is made to rotate and drags the sphere on a circular path on the glass slide, while measuring both the normal force and rotational torque simultaneously. The moment arm of the frictional force is 7 mm, which is much larger than a typical contact circle ($<120$ $\mu$m  in diameter in normal forces less than 100 mN) and the sliding distance in our experiments. The rotational inertia of the whole rotating part including the sample holder and the sphere is measured prior to setting up the contact and subtracted by the rheometer, so that the torque is zero in non-contact condition. Therefore, the recorded torque directly gives the friction at the sphere and glass interface.

A typical friction experiment [pink curve, Fig. 1(a)] shows that the frictional force is highest at the onset of slip, and then decreases to its steady dynamic value. This is the usual observation that static friction coefficient is larger than the dynamic one. The transition, however, is continuous and extends over a longer time at a smaller sliding velocity (red curve). The uniqueness of our setup is that it can reliably go to very low rotation rates. We find that if the sliding velocity decreases to very small values ($\sim$ nm/s), the frictional force, starts from zero and slowly increases to an apparent steady value [Fig. 1(b)]. Such transients with both increasing and decreasing friction have been observed for a variety of sliding materials; notably in slow frictional sliding of rocks in laboratory studies of tectonics and earthquake dynamics \cite{11, 22}. They are commonly described in the phenomenological rate and state friction framework, in which the frictional strength of the contact can show either of slip weakening or strengthening, depending on the slip velocity. Our sliding experiments reveal another distinct behavior of a frictional interface: in any sliding velocity, the transient dynamics fade away when the experiment is repeated with the same velocity immediately afterwards [grey curves, Figs. 1(a) and 1(b); and Supplemental Material, Fig. S1 for polypropylene spheres]. Only after significant waiting (aging) or refreshing the contact, the pronounced transients reappear again. This clearly indicates that the state of the interface evolves during the first sliding and the interface ‘remembers’ its sliding history in the next run. The repeated runs show a characteristic linear force which starts from zero and increases up to the plateau friction, where sliding occurs. This linear force is due to the elastic response of the measurement system, which we can subtract from the first-run sliding dynamics [Fig. 1(b), blue curve] to obtain the actual frictional strength (purple curve, again increasing from zero). Indeed, a part of the rotational deflection of the rheometer, which is advancing with a constant rate, is accommodated by the elastic deformation of the sphere and the rheometer components (instead of sliding at the interface). This elastic contribution can be easily separated from the first friction dynamics of the interface at low sliding speeds, by making the replacement:
\begin{equation}
t-\frac{F(t)}{kv} \rightarrow t_{corr}.
\end{equation}
Where $kv$, with $k$ the elastic shear stiffness of the total measurement system (rheometer + elastic deformation of the bead) and $v$ the sliding velocity, is the slope of the red dashed line in Fig. 1(b). This therefore removes the time needed for the rheometer to store the elastic force of $F$ in the system from the original data, as also shown by black arrows in the figure. In the case of infinite stiffness of the system, $k\rightarrow \infty$ , or very large sliding speeds, $v\rightarrow \infty$, no correction is needed.

\begin{figure}
	\begin{center}
		\includegraphics[scale=0.65]{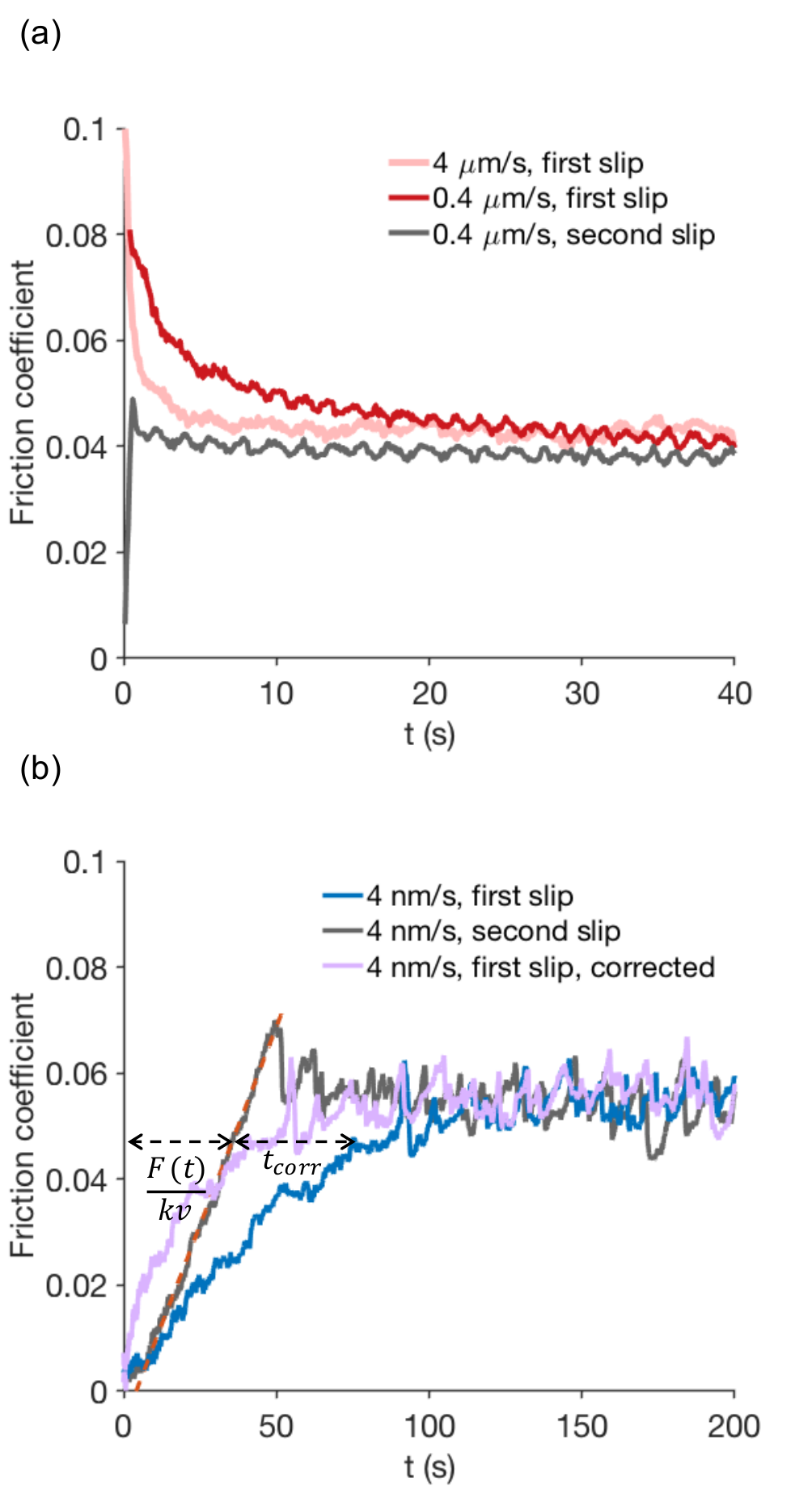}
	\end{center}
	\vspace{-.5cm}
	\caption{Transient friction at the onset of sliding for PTFE sphere-on-glass frictional system. (a) At high sliding velocities friction starts off from a point above the steady-state value. It then gradually decays to the steady state, passing through a slow transient. The second run of the same interfacial system (grey curve, following after the red curve) does not show the pronounced transient anymore. (b) At low sliding velocities, friction increases from zero to an apparent plateau value. Again, the slow transient behavior vanishes in the second slip. Instead, a linear force, due to the elastic response of the system, builds up until the interface slides at the plateau friction. We subtracted this elastic reaction from the first sliding force to obtain the actual friction dynamics (purple curve). In fact, the recorded time is replaced with the horizontal (time) distance between the initial dynamic and the elastic characteristic line (black arrows).}
	\label{f:numeric}\end{figure}

When corrected for the elastic response of the measurement system, the deceasing and increasing friction transients in high and low sliding speeds persist, and we now ask how these are related? To investigate this question, we observed the dynamics of the frictional interface in a slow sliding experiment over a much longer time (typically a few hours), corresponding to a sliding distance comparable to the fast sliding case ($\sim$ 25 $\mu$m). Figure 2 shows the evolution of the frictional strength of the interface when it slides with speed of 1 nm/s in an isolated noise-free environment. The linear elastic contribution has been subtracted from the plot. Remarkably, the interface frictional strength, after increasing for a long time ($\sim$ 1000 s), reaches a peak value and then decays very slowly to its steady-state (dynamic) friction. This non-monotonic behavior cannot be described by the standard rate-and-state theory. The effect is not simply due to wear and/or plastic changes of the surface roughness, since it reappears each time the contact is renewed, independent of how many times two surfaces have previously moved over each other. It is inherent to the new interfacial system and results from the evolution of the contact area itself by aging and shearing. 

\begin{figure}[hbt]
	\begin{center}
		\includegraphics[scale=1]{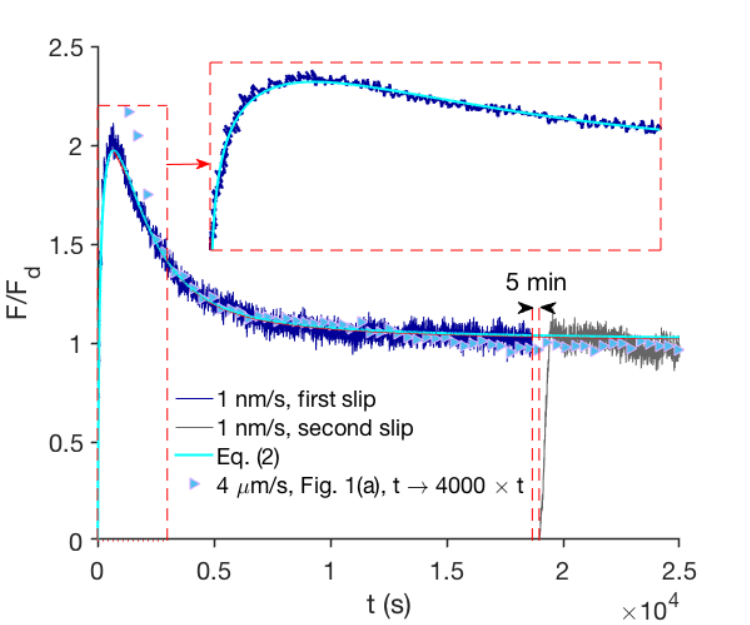}
	\end{center}
	\vspace{-.5cm}
	\caption{Non-monotonic friction dynamics for sliding of a typical PTFE sphere on glass surface. The sliding speed is 1 nm/s. It takes approximately 5 hours for the frictional system to reach the steady-state dynamic friction. The elastic response of the measurement system has been subtracted. A second sliding run, started 5 minutes after the first run, is also shown (grey line). Cyan solid line is Eq. (3), with $\alpha=0.92$ $\mu m^{-1}$, $a/F_{d}=0.62$ and $t_{0}=14.8$ s as fitting parameters. The inset shows the timeframe from 0 to 3000 s. Triangles are data from Fig. 1(a), 4 $\mu$m/s, when the time axis is extended 4000 times. This implies that, although the two transient dynamics occur on very different time scales (2-3 s vs 25,000 s), they are similar when plotted as a function of sliding distance.}
	\label{f:numeric}\end{figure}

In order to understand the friction transients, we consider that for a stationary interface, logarithmic aging implies $dF/dt=a/t$, where $t$ is the age of the interface and $a$ is a constant. We assume a ‘shear rejuvenation’ process that counteracts the aging \cite{23,24}; the simple assumption is that the rejuvenation rate is proportional to the sliding velocity, $v$, and how far the system is from its steady-state dynamic friction, $F(t)-F_{d}$ \cite{24}. The competition between aging and shear-induced rejuvenation of the asperity contacts leads to a simple differential equation for the friction force:
\begin{equation}
\frac{dF}{dt}=\frac{a}{t}-\alpha v (F-F_{d}).
\end{equation}
The constant $\alpha$, with dimension of m$^{-1}$, contains the shearing characteristics of the contact interface. Dividing both sides of Eq. (2) by $v$  reveals that the assumption of shear-induced rejuvenation being proportional to the sliding rate is equivalent to a simple fundamental statement about sliding friction: the state of the sliding interface depends solely on the total sliding distance $x=vt$, as we indeed see in the low-velocity dynamics [Fig. (3)]. This has also been inspected already by extensive studies of transient phenomena in rocks friction (see \cite{11} and references in there). In the rate and state formulations, a characteristic sliding displacement is introduced to describe the evolution dynamics after a change of sliding conditions \cite{11,25,26}. It has been shown that this distance is independent of sliding speed and normal stress, but changes with, e.g. surface roughness. Moreover, the fact that the repeated sliding runs (grey curves, Figs. 1 and 2) do not demonstrate the transient friction peak anymore means that the evolution of the system in time, perhaps due to plastic flow at asperities, is negligibly slower than the sliding evolution imposed by Eq. (2). If we stop sliding, the only thing that changes is the bulk elastic energy of the system. As soon as this elastic deformation is restored again by the rheometer (the linear force), the interface slips with the same frictional force as when it was stopped. 

\begin{figure}[hbt]
	\begin{center}
		\includegraphics[scale=0.65]{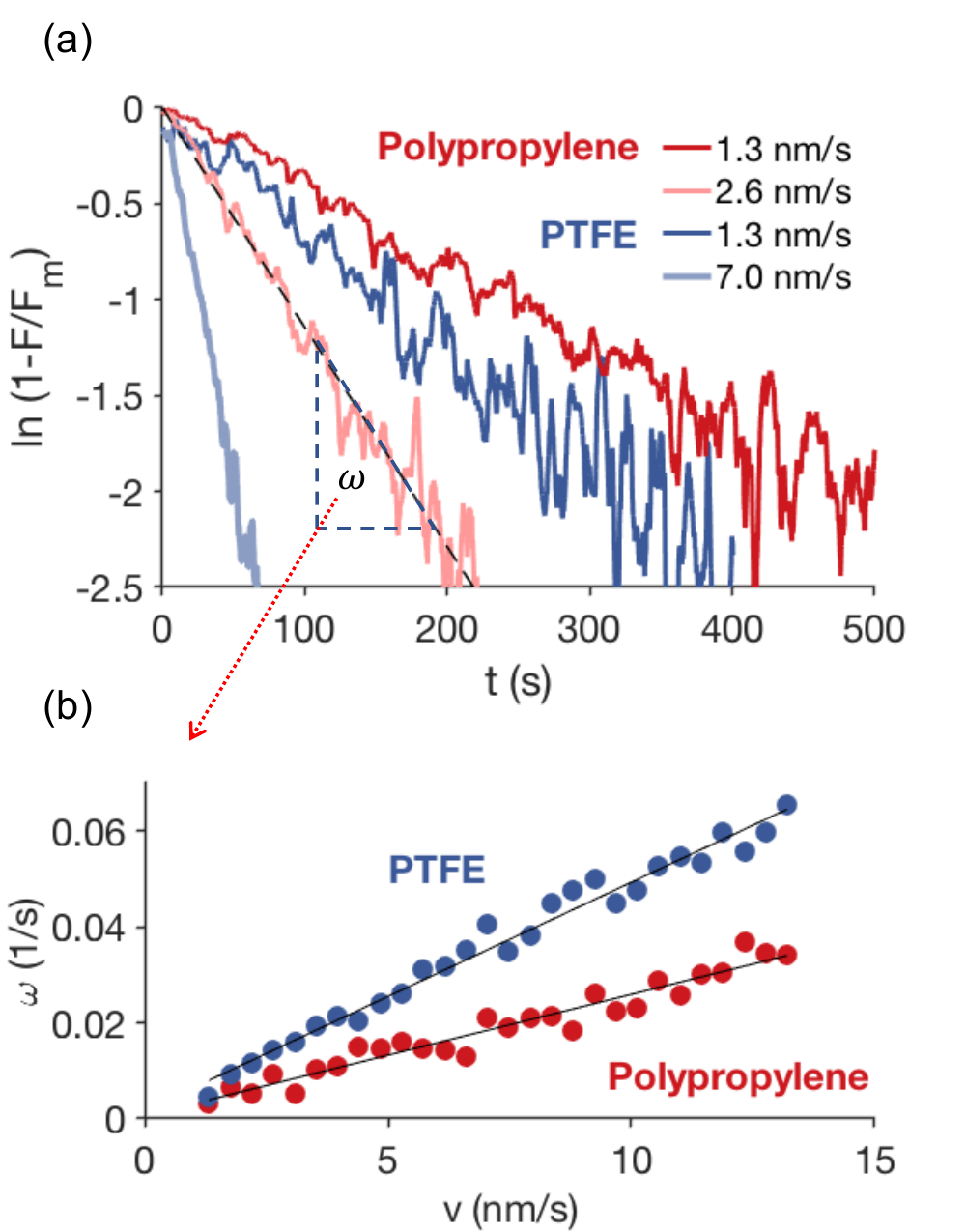}
	\end{center}
	\vspace{-.5cm}
	\caption{The increasing slow-sliding dynamics, before subtracting the elastic contribution, can be approximated by an exponential fit: $ln(1-F/F_{m})=\omega t$, with $\omega$ linearly proportional to $v$ so that $ln(1-F/F_{m}) \sim v t$. $F_{m}$ is the maximum friction in each case. Since the elastic force is also proportional to shear deflection $vt$, the actual friction evolution derived from their difference must also be a function of the total sliding distance $vt$, in agreement with Eq. (2).}
	\label{f:numeric}\end{figure}

Solutions of Eq. (2) are of the form
\begin{equation}
F(t)=a e^{-\alpha v t} \int_{t_{0}\rightarrow 0}^t \frac{e^{\alpha v t'}}{t'}dt' +F_{d}(1-e^{-\alpha v t}),
\end{equation}
where $t_{0}\rightarrow 0$ is a singular point in time from which the interfacial system evolves by aging and rejuvenation. Such a singular behavior arises in all disordered systems demonstrating logarithmic aging, ranging from creased/crumpled paper \cite{27,28} and granular piles \cite{29} to static frictional interfaces \cite{7,12,30}. The consequence is that, if $t_{0}\rightarrow 0$, the first term of Eq. (3) yields infinite forces, which is obviously not observed in reality. For friction to be finite, there is a cut-off at short times, determined by how the two materials were brought in contact with each other \cite{30}. From this initial state, the sliding process, collectively described by the above evolution equation, starts. We find this simple description to work even quantitatively: Equation (3) can reproduce the observed non-monotonic friction evolution extremely well (Fig. 2, and Supplemental Material, Fig. S2). Imposing a small sliding speed, the friction force first increases and goes to values above the steady-state friction by the integral originating from the aging term. It subsequently decays exponentially due to the rejuvenation. Figure 2 also shows that, although the 4 $\mu$m/s transient dynamics of Fig. 1(a) extends over a much shorter time scale, it is similar to the much longer 1 nm/s sliding dynamics when plotted as a function of the sliding distance. 

The above understanding of friction provides a comprehensive picture of how and when a frictional slip occurs, and in addition explains why it is generally very difficult to reproduce the static friction coefficient in experiments \cite{5,6,23}. While rejuvenation or de-aging of the asperity contact area starts with the smallest applied tangential force \cite{14}, a macroscopic slip is only triggered and maintained if the applied shear force is larger than the peak frictional strength of Eq. (3). However, both the height and the time of this peak are determined by the aging and shear dynamics, given by Eq. (3); and in this way, they depend on the history and initial state of the interface. Consequently, the threshold shear force of the onset of slip cannot be defined as a simple material constant, i.e. the static friction coefficient \cite{5}. In fact, due to the singular nature of aging, the non-monotonic dynamics are very sensitive to the initial state of the interface, represented by $t_{0}$. Figure 4 shows how our modelling predicts that the peak friction value, and hence the static friction coefficient to be a logarithmic function of $t_{0}$. Small changes in $t_{0}$ that are experimentally hard to control lead to significant variations in the static friction.  This description of the onset of sliding friction also gives an explanation for why micrometer-scale motions can make large changes in the frictional properties of an interface \cite{23}, corresponding to ubiquitous observations that small pre-vibrations can help to overcome the initial static friction when one is sliding a heavy object over the floor. The partial de-aging (rejuvenation) imposed by these small vibrations can change the dynamic pathway of the interface, leading to a smaller static frictional barrier.

Our results yield a fundamentally new understanding of the onset of slip at frictional interfaces. The simple, but remarkably non-monotonic, friction evolution shows that the classical concept of the static friction is ill-defined and the static friction barrier in the transition from static to steady-state sliding of a frictional interface emerges through the continuous dynamics of the onset of slip. This should be valid for describing the slip behavior of all frictional systems that undergo aging, including slow slip processes along an earthquake-generating fault. Analyzing and understanding these processes is critically important for monitoring the nucleation of earthquakes  \cite{25,26,31}.

\begin{figure}[hbt]
	\begin{center}
		\includegraphics[scale=1]{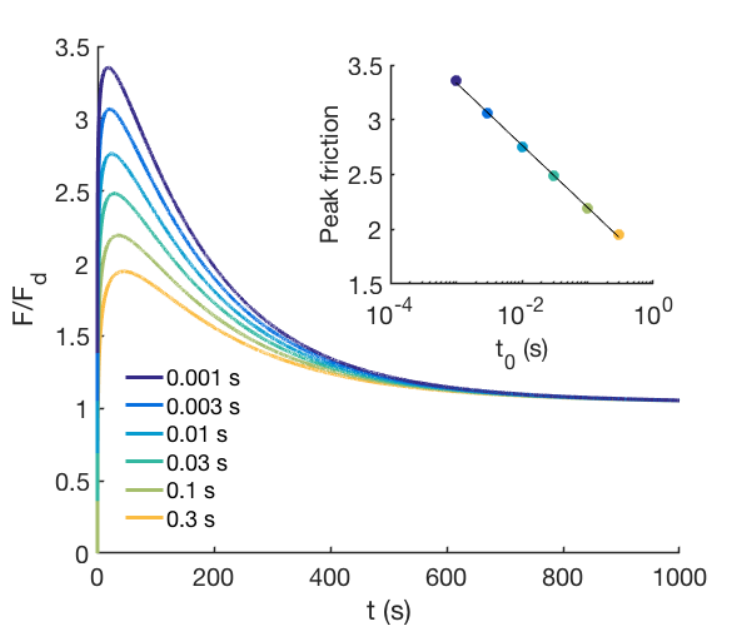}
	\end{center}
	\vspace{-.5cm}
	\caption{Computed friction evolution for a system with $\alpha v=0.007$ $s^{-1}$ and $a/F_{d}=0.3$  for different values of $t_{0}$  from 0.001 to 0.3 s. Because of the singular behavior of frictional aging, the peak value or static friction depends logarithmically on $t_{0}$  (inset). }
	\label{f:numeric}\end{figure}

\vspace*{0.3 cm}
This project has received funding from the European Research Council (ERC) under the European Union’s Horizon 2020 research and innovation program (Grant agreement No. 833240)

\end{document}